\documentclass[twocolumn,showpacs,preprintnumbers,amsmath,amssymb,nofootinbib]{revtex4}

\usepackage[latin1]{inputenc}
\usepackage{pslatex}
\usepackage{nicefrac}
\usepackage{wasysym}
\usepackage{epsfig}

\begin{document}

\title{Phantom Accretion by Black Holes and the Generalized Second Law of Thermodynamics}

\author{J. A. S. Lima} \email{limajas@astro.iag.usp.br}

\author{S. H. Pereira} \email{spereira@astro.iag.usp.br}

\author{J. E. Horvath} \email{foton@astro.iag.usp.br}

\affiliation{Universidade de S\~ao Paulo -- Instituto de Astronomia,
  Geof\'\i sica e Ci\^encias Atmosf\'ericas \\
Rua do Mat\~ao, 1226 -- 05508-090 Cidade Universit\'aria, S\~ao Paulo, SP,
Brazil}

\author{Daniel C. Guariento} \email{carrasco@fma.if.usp.br}

\affiliation{Universidade de S\~ao Paulo -- Instituto de F\'\i sica\\
Rua do Mat\~ao, Travessa R, 187 -- 05508-090 Cidade Universit\'aria, S\~ao
Paulo, SP, Brazil}

%\keywords{Dark energy, cosmic distance, Inhomogeneity Parameter}

\begin{abstract}

The accretion of a phantom fluid with non-zero chemical potential
by black holes is discussed with basis on the Generalized Second
Law of thermodynamics. For phantom fluids with positive
temperature and negative chemical potential we demonstrate that
the accretion process is possible, and that the condition
guaranteeing the positiveness of the phantom fluid entropy
coincides with the one required by Generalized Second Law. In
particular, this result provides a  complementary confirmation
that cosmological phantom fluids do not need to have negative
temperatures.

\end{abstract}

\maketitle

\section{Introduction}

The discovery of the present accelerating stage of the Universe
from Supernovae (SNe) type Ia observations \cite{SN} is one of the
most important achievements in modern cosmology. This result has
been indirectly confirmed by a large set of independent
complementary observations, involving the temperature anisotropies
of the cosmic microwave background \cite{CMB}, the X-ray surface
brightness from galaxy clusters \cite{Clusters} and the age of the
universe from globular clusters and old high redshift galaxies
\cite{Old}. In the context of general relativity, an accelerating
stage can be explained by assuming the existence of an exotic
substance with negative pressure dubbed dark energy (or
quintessence) which represents an intriguing evidence for physics
beyond the domain of the standard  model of particle physics.

Theoretically, there are many candidates to represent this extra
non-luminous (in addition to cold dark matter) relativistic
component \cite{review}. Many authors have worked on the idea that
the unknown, uncoupled dark energy component is due exclusively to
a minimally coupled scalar field  (quintessence field) which has
not yet reached its ground state and whose current dynamics is
basically determined by its potential energy \cite{Quint}. A
special attention has also been paid to the so-called XCDM
cosmologies which are phenomenologically  described by an equation
of state (EoS) of the form \cite{XCDM}
\begin{equation}\label{eqstate}
p=\omega \rho,
\end{equation}
where $p$ and $\rho$ denotes  the pressure and energy density,
respectively, and  $\omega$ is a constant negative parameter. The
case $\omega = -1$ corresponds to a positive cosmological
constant, or vacuum energy, while for $\omega < -1$ we have the so
called phantom dark energy regime or phantom fluids. Depending on
the form of potential, this kind of fluid can also be represented
by a scalar field model \cite{phantom}.

It is widely known that SNe data alone rules out non-accelerating
models with very high significance level, and that even dark
energy candidates with $\omega > -2/3$ are strongly disfavored.
Actually, some analyzes based on the combination of SNe and large
scale structure data imply that the EoS parameter lies within the
range $-1.48 < \omega < -0.72$ at 95\% confidence level
\cite{tonry}.  Therefore, the possibility of a phantom regime
($\omega < -1$) cannot be excluded.

The study of the phantom regime present some peculiar features,
for example, stability issues \cite{pg11,pg12}, increasing energy
density with cosmic time, the possibility of superluminal sound
speed \cite{pg4} and even the violation of the dominant energy
condition is possible (as the phantom fluid must satisfy $\rho + p
< 0$). The dominance of the cosmological content by a phantom
fluid implies that the universe will undergo a very fast expansion
in the future when it finally reaches the big-rip singularity.

Another intriguing point is related to the thermodynamic
properties underlying the phantom energy fluids. Some earlier
studies developed by Lima and coworkers \cite{limamaia,LA04} shown
that the temperature of the phantom fluid without chemical
potential is positive definite but its co-moving entropy,
$S_{\mathrm{ph}} < 0$,  is always negative (for a different
analysis showing the same result see the paper by Brevik {\it et.
al}. \cite{Brevik}). Therefore, unless  one disregards the
statistical microscopic concept of entropy from which $S>0$, the
phantom fluid cannot exist in nature (see also \cite{gonzalez} for
a different approach based on the idea that the temperature of
phantom fields  is negative).

Later on, Izquierdo and Pav\'on \cite{pavon} discussed the phantom
energy in the context of the generalized second law (GSL) of
thermodynamics. It was found that although the entropy constraint
be violated, the GSL may be satisfied, that is, ${\dot S} = {\dot
S}_{\mathrm{ph}} + {\dot S}_{\mathrm{h}} \geq 0$, where
$S_{\mathrm{h}}$ is the horizon entropy.

More recently, the thermodynamic and statistical properties of
phantom fluids were reexamined by considering the existence of a
non-zero chemical potential ($\mu$). In this case, it was found
that the entropy condition, $S \geq 0$, implies that the possible
values of $\omega$ are heavily dependent on the value, as well as
on the sign of  the chemical potential \cite{limasaulo1} (for a
related study in K-essence dark energy models see the paper by
Bili\'c \cite{Bilic}).  For $\mu >0$, the $\omega$-parameter must
be strictly greater than -1 (vacuum is forbidden) while for $\mu <
0$ not only the vacuum but even a phantom-like behavior ($\omega
<-1$) is allowed. In any case, the ratio between the chemical
potential and temperature remains constant, that is,
$\mu/T=\mu_0/T_0$, where $\mu_0$, $T_0$ are the present day values
of the chemical potential and temperature, respectively. Still
more  important, the positiveness of the entropy is preserved for
all kinds of dark energy components as long as the condition

\begin{equation}\label{cond}
\omega \geq -1 + \frac{\mu_0 n_0}{\rho_0},
\end{equation}
is satisfied, where  $n_0$ and $\rho_0$ are, respectively, the
present values of the particle number density and energy density
of the dark energy fluid described by Eq. (1). Actually, if the
chemical potential is negative, the above expression shows that
the $\omega$ parameter is effectively describing a phantom fluid.

In this context, it is interesting to investigate  the constraints
associated to the validity of GSL and the possibility of a
positive critical mass enabling the phantom accretion onto black
holes. As we shall see, the positiveness of this mass leads to a
remarkable result, namely: the same condition for the $\omega$
parameter  given above is also required by GSL. In particular,
this means that cosmological phantom fluids do not need to have
negative temperatures. The present study is complemented with some
numerical estimates of the critical mass based on the Holographic
Bound, that is, the idea that there is an upper limit to the
entropy inside the "surface" limiting the system.

\section{Accretion process of phantom fluids}

Following the results presented in \cite{limasaulo1}, let us now
discuss the thermodynamic aspects governing the accretion of of
phantom fluids by  black holes when the chemical potential of the
phantom fluid is non zero. The main results obtained here will be
compared with the recent analysis by Pacheco and Horvath
\cite{jorge} in the absence of chemical potential.

The change on the black hole mass due to the accretion process in
a universe filled with a dark energy fluid was first investigated
by Babichev, Dokuchaev and Eroshenko \cite{babichev}. In the case
of phantom fluids, they  found that the black hole mass changes at
a negative rate, that is, $\dot{M} < 0$.  Mathematically, such a
behavior can be readily understood. In a general accreting process
the black hole mass rate is proportional to the enthalpy density
of the fluid ($\rho$ + p). Therefore, when the  dominant energy
condition (DEC) is obeyed (violated), the black hole mass will
increase (decrease) during the accretion process.  In this second
case, the immediate consequence is that the area of the black hole
horizon will decrease along with its entropy ($S \propto M^{2}$,
where M is the mass of the Schwarzschild black hole).

On the other hand, Gao and collaborators \cite{gao} claimed that
the back-reaction provoked by the accretion of phantom matter on
the black hole metric was ignored in the mentioned calculations.
In a low matter density background, this effect can be safely
ignored, but if the background matter density is large (comparable
to the black hole density), the metric describing this black hole
will be significantly modified, and the back-reaction must
necessarily be considered. Different from  the authors of
\cite{babichev} they found that the physical black hole mass will
not decrease but rather grows due to the accretion of phantom
fluid, so that $\dot{M} > 0$. However, Mart\'\i n-Moruno \emph{et
al.} \cite{moruno} have argued that the Gao {\it et al}. findings
are misleading because it was assumed that the black  hole mass
function is proportional to the scale factor. In particular, the
entropy of the fluid has not been taken into account in the
solution for the accretion process, and, as such, the result seems
to be based on a somewhat incomplete picture.  For completeness
and by considering  that the results are still under debate, in
what follows we will discuss the equilibrium conditions for both
cases ($\dot{M} > 0$ and $\dot{M} < 0$).

\subsection{Phantom Accretion by Black Holes: GSL Constraints}

In terms of the present day quantities, the energy density of a
dark fluid can be written as \cite{limamaia,LA04,Brevik}

\begin{equation}\label{energ}
\rho = \rho_0 \left(\frac{T}{T_0} \right)^{\frac{1 + \omega}{\omega}}\,,
\end{equation}
whereas its entropy (including a chemical potential) reads \cite{limasaulo1}

\begin{equation}\label{entropy}
S = \left[\frac{(1 + \omega) \rho_0 - \mu_0 n_0}{T_0}\right]
\left(\frac{T}{T_0}\right)^{\frac{1}{\omega}}\,.
\end{equation}
where $T$ is the temperature ($T_0$ is its present day value). It
thus follows that the total entropy of the system consisting of a
(Schwarzschild) black hole plus a dark energy fluid reads (unless
explicitly stated, in our units $\hbar = k_{B} = c=1$),

\begin{equation}\label{entropyS}
S = {4 \pi G M^2} + \left[\frac{(1 + \omega) \rho_0 - \mu_0
    n_0}{T_0}\right] \left(\frac{\rho}{\rho_0}\right)^{\frac{1}{1 +
    \omega}} V\,,
\end{equation}
where the first term represents the black hole entropy and the
second term is the phantom fluid entropy inside a co-moving volume
$V$, written in terms of the energy density. Now, due to the
accretion process, in an arbitrarily short time interval, the
black hole mass varies by $\Delta M$ and the phantom field energy
varies by $\Delta \rho$. Therefore, the total entropy variation
within the cavity takes the form

\begin{widetext}
\begin{equation}\label{bhm}
\Delta S = {8\pi G M} \Delta M + \frac{1}{(1 + \omega)}
\left[\frac{(1 + \omega) \rho_0 - \mu_0 n_0}{T_0 \rho_0}\right]
\left(\frac{\rho}{\rho_0}\right)^{-\frac{\omega}{1 + \omega}} V \Delta \rho\,.
\end{equation}
\end{widetext}
For a phantom fluid modelled by a scalar field, only the kinetic\\
term contributes to the accretion so that the energy conservation
inside the cavity implies that

\begin{equation}\label{deltam}
\Delta M = - \frac{1}{2} (1 + \omega) V \Delta \rho \,.
\end{equation}

Now, by inserting  (\ref{deltam}) into (\ref{bhm}), we obtain an
expression for the total entropy variation of the black hole plus
dark energy:

\begin{widetext}
\begin{equation}\label{bh}
\Delta S = \left\{{8\pi G M} - \frac{2}{(1 + \omega)^2}
\left[\frac{(1 + \omega) \rho_0 - \mu_0 n_0}{T_0 \rho_0}\right]
\left(\frac{\rho}{\rho_0}\right)^{-\frac{\omega}{1 + \omega}}\right\} \Delta M\,.
\end{equation}
\end{widetext}

We move now to analyze the two cases previously presented.
According to Gao \emph{et al.} \cite{gao}, the mass of the black
hole increases with time during the  phantom accretion process,
that is, $\Delta M \geq 0$. In order to satisfy the GSL ($\Delta S
\geq 0$), we must have the term inside the curly brackets
appearing in \eqref{bh} positive definite thereby yielding the
condition:

\begin{equation}\label{mcrit}
M \geq M_{\mathrm{crit}} = \frac{1}{4\pi G(1 +
  \omega)^2} \left[(1 + \omega) - \frac{\mu_0 n_0}{\rho_0}\right] \frac{1}{T_0}
\left(\frac{\rho}{\rho_0}\right)^{-\frac{\omega}{1 + \omega}}\,.
\end{equation}

Note that for $\mu_{0} = 0$ the above critical mass reduces to the
one derived in Ref. \cite{jorge} (see their equation (14)). In
this case, the critical mass is negative when $\omega<-1$ and the
inevitable conclusion is that the process is physically forbidden.
In the point of view of  Refs. \cite{Brevik,LA04}, such a result
should be physically expected since phantom fluids with negative
temperature cannot exist in nature. On the other hand, we see that
for negative values of $\mu_0$ there exists a {\it positive}
critical mass above which the black hole can accrete the phantom
fluid.

If we adopt the results from Babichev {\it et al}. \cite{babichev},
where $\Delta M < 0$, the condition to the mass is $M \leq
M_{\mathrm{crit}}$. Therefore, only black holes with mass \emph{below} the
critical mass can accrete phantom energy.

An interesting result emerging from the above discussion based on
the validity of GSL is that to keep the positiveness of the
critical mass, the following condition must be  satisfied (the
term outside the square brackets in \eqref{mcrit} is positive
definite and thus the term inside the square brackets must be
positive as well)

\begin{equation}\label{condGSL}
\omega \geq -1 + \frac{\mu_0 n_0}{\rho_0},
\end{equation}
which is just condition \eqref{cond} previously obtained in
\cite{limasaulo1} from a thermodynamical analysis based
exclusively on the phantom fluid thermodynamic properties. The
fact that the same condition emerges naturally from the GSL
constraints reinforces the claims that phantom fluids must have
positive temperatures.

\subsection{Critical Mass Estimates}

The above constraint has an additional interesting property. It
can be used to give an upper bound estimate to the present day
value of the chemical potential, more precisely to the product
$\mu_0 n_0$. Notice that the inequality \eqref{cond} can be
written as

\begin{equation}\label{cond2}
\mu_0 \leq \frac{\rho_0}{n_0} (1 + \omega)\,.
\end{equation}
It should also be recalled that  the condition \eqref{cond} for
the chemical potential is a general result, and, as such, it is
not related to our previous analysis of phantom fluid accretion by
black holes with basis on GSL. Since the chemical potential has
the dimension of energy per particle, the product $\mu_0 n_0$ has
the dimension of energy density, like $\rho_0$. Assuming that our
universe is flat, the critical density is given by $\rho_c \equiv
\nicefrac{3H_0^2}{8\pi G} \simeq 4 \times 10^{-47}$~GeV$^4$.
%\rho_c \equiv \nicefrac{3H_0^2}{8\pi G} \simeq 1.24 \times
%10^{-43}$~$\nicefrac{\mathrm{J}}{\mathrm{m}^3}
If the dark energy contributes with about 70\% of the matter content
of the universe, we may suppose that the present day value of the dark
energy density is given by $\rho_0 \simeq 0.7 \times \rho_c \simeq 2.8
\times 10^{-47}$~GeV$^4$.
%\rho_0 \simeq 0.7 \times \rho_c \simeq 8.70 \times
%10^{-44}$~$\nicefrac{\mathrm{J}}{\mathrm{m}^3}
This allows us to estimate

\begin{equation}
\mu_0 n_0 \leq (1 + \omega) \rho_c\,.
\end{equation}

Taking as an example the value $\omega = -3/2$, we have \mbox{$\mu_0 n_0 \leq
-1.4\times 10^{-47}$~GeV$^4$}.
%$\mu_0 n_0 \leq -4,35 \times
%10^{-44}$~$\nicefrac{\mathrm{J}}{\mathrm{m}^3}$.

In order to study some limits of \eqref{mcrit}, let us define \mbox{$\alpha
\equiv -\nicefrac{\mu_0 n_0}{\rho_0}$} the ratio between the chemical
potential density and the energy density, which is a constant
dependent only on present day values. Using the relation
\eqref{energ}, we rewrite the critical mass \eqref{mcrit} as

\begin{equation}
M_{\mathrm{crit}} = 4.5 \times 10^{-20}
%4.90 \times 10^{-7}
\left[\alpha - \frac{1}{2}\right] \frac{M_{\astrosun}}{T_{GeV}}\,,
\end{equation}

\noindent
where $M_{\astrosun}$ is the solar mass, $T$ is the temperature of the
phantom fluid, given in GeV, and $\omega = -\nicefrac{3}{2}$ is used. This
shows that the critical mass is proportional to the inverse of the
temperature of the phantom fluid and also depends on $\alpha$.

\section{Critical Mass from Holographic Principle}

A more quantitative analysis of the critical mass value requires
knowledge of its entropy, and consequently of its chemical
potential, which remains indeterminate since the remaining
constants \,$n_0,\,\rho_0$ and $T_0$ are unknown. However, an
upper bound can be derived from Holographic Principle bound which
states that

\begin{equation}
{4\pi\over 3}R^3 s \leq {\pi\over l_P^2}R^2\,,
\end{equation}

\noindent where $R$ is the event horizon radius, $s$ is the
entropy density and $l_P$ is the Planck length. It can be showed
that, in a phantom dominated universe, the Holographic Bound is
satisfied if (see \cite{jorge} and references therein)

\begin{equation}
s_0\leq {3|1+3\omega|\over 8}{\sqrt{\Omega_{DE}}H_0\over
l_P^2}\simeq 7.2\times 10^{-5}|1+3\omega|\textrm{GeV}^{-3}\,.
\end{equation}

\noindent
In our case, using \eqref{entropy}, we have

\begin{equation}
-\mu_0 n_0\leq -(1+\omega)\rho_0+T_0 {3|1+3\omega|\over
8}{\sqrt{\Omega_{DE}}H_0\over l_P^2}\,,
\end{equation}

\noindent
or in terms of the $\alpha$ parameter defined above

\begin{equation}
\alpha \leq {1\over 2}+3.9\times 10^{42} T_0
\end{equation}
where we have used $\omega=-3/2$ and $\rho_0=0.7\times \rho_{crit}$, and $T_0$ is given in GeV.
\begin{figure}[t]
\begin{center}
\epsfig{file=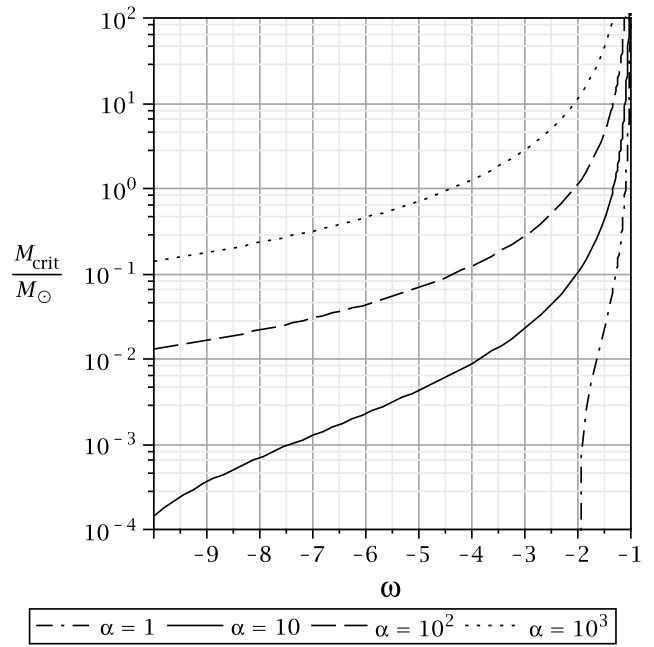, scale=0.45}
\end{center}
\caption{The present Black hole critical mass as a function of
$\omega$ for some selected values of  $\alpha$. }\label{fig1}
\end{figure}

Now, by considering  the case of very small temperature, $\alpha
\approx 1/2$ if we take $T_0\simeq 0$, and thus $M_{\mathrm{crit}}
\simeq 0$. This shows that all black holes can accrete in the case
$\Delta M \geq 0$, but the process is forbidden for case $\Delta M
\leq 0$.

As discussed by several authors, the thermodynamic treatment of
dark energy fluids implies that its temperature increases in the
course of the expansion. Thus, a more interesting case occurs if
we take, for instance, $T_0\simeq 10^{-19}$GeV $\simeq 10^{-6}$K,
and so $\alpha \approx 10^{24}$. We have the critical mass
$M_{\mathrm{crit}} \simeq \nicefrac{4.6\times
10^{3}M_{\astrosun}}{T}$. Today this critical mass is
$M_{\mathrm{crit}} \simeq  4.6\times 10^{22}M_{\astrosun}$, a very
large mass, but of the same order of that obtained in \cite{jorge}
by a different method.

A more realistic case occur if we take $\alpha \approx 10$. In
this case the critical mass today is $M_{\mathrm{crit}} \simeq 2.2
M_{\astrosun}$, the order of a solar mass. It is interesting to
note the dependence of the critical mass with $\omega$ for some
values of $\alpha$. It is plotted in Fig. 1.

\section{Conclusions}\label{sec-concl}

In this work we have discussed the accretion of phantom fluids
with negative chemical potential by black holes. As we have seen,
there is a positive critical mass in order to enable the phantom
accretion. Some numerical estimates constraining  the critical
mass threshold based on the adoption of reasonable values of the
chemical potential, or more directly, in terms of  the Holographic
Bound, have also be presented, even if the latter has not been
rigourously proved to hold for arbitrary space-times.

As physically expected, the Generalized Second Law of
thermodynamics determines the thermodynamic viability of the whole
process and the amount of dark energy accretion. Depending on the
kind of analysis adopted, the process may either increase or
decrease the mass range of black holes accreting the phantom fluid
along the history of the universe. More interesting, the
positiveness of the critical mass (based on GSL) leads to the same
constraint involving the present day values of the physical
quantities specifying the phantom fluid and previously obtained
through  a thermodynamic analysis of the phantom energy alone
\cite{limasaulo1}. In other words, phantom fluids with zero
chemical potential are not consistent because they require either
a negative entropy (which is microscopically unacceptable) or a
negative temperature (which needs a bounded spectrum which has not
been justified from any scalar field model).

\begin{acknowledgments}

The authors would like to thank V. C. Busti, J. V. Cunha, J. F. Jesus,
A. C. Guimar\~aes, R. Holanda, R. C. Santos and P. I. Braun for helpful
discussions. JASL is partially supported by CNPq and FAPESP
No. 04/13668-0 and SHP is supported by CNPq No. 150920/2007-5
(Brazilian Research Agencies). DCG and JEH are supported by CNPq
through grants and fellowships.

\end{acknowledgments}

\end{document}